\documentclass[aps,prd,groupedaddress,amsmath,amssymb,showpacs]{revtex4}
\usepackage{graphicx}
\usepackage{dcolumn}
\usepackage{bm}
\begin{document}
\def\d{{\rm d}}
\def\lp{\left. }
\def\rp{\right. }
\def\lr{\left( }
\def\rr{\right) }
\def\le{\left[ }
\def\re{\right] }
\def\lg{\left\{ }
\def\rg{\right\} }
\def\lb{\left| }
\def\rb{\right| }
\def\beq{\begin{equation}}
\def\eeq{\end{equation}}
\def\bea{\begin{eqnarray}}
\def\eea{\end{eqnarray}}
\def\as{\alpha_s}
\def\nbar{\bar{N}}
\def\bbar{\bar{b}}
\def\nn{\nonumber}

\title{Joint resummation for slepton pair production at hadron
colliders}
\author{Giuseppe Bozzi}
\affiliation{Insitut f\"ur Theoretische Physik, Universit\"at
Karlsruhe, Postfach 6980, D-76128 Karlsruhe, Germany}
\author{Benjamin Fuks}
\author{Michael Klasen}
\email[]{klasen@lpsc.in2p3.fr} \affiliation{Laboratoire de
Physique Subatomique et de Cosmologie, Universit\'e Joseph
Fourier/CNRS-IN2P3, 53 Avenue des Martyrs, F-38026 Grenoble,
France}
\date{\today}
\begin{abstract}
We present a precision calculation of the transverse-momentum and
invariant-mass distributions for supersymmetric particle pair
production at hadron colliders, focusing on Drell-Yan like slepton
pair and slepton-sneutrino associated production at the CERN Large
Hadron Collider. We implement the joint resummation formalism at
the next-to-leading logarithmic accuracy with a
process-independent Sudakov form factor, thus ensuring a universal
description of soft-gluon emission, and consistently match the
obtained result with the pure perturbative result at the first
order in the strong coupling constant, i.e.\ at
$\mathcal{O}(\as)$. We also implement three different recent
parameterizations of non-perturbative effects. Numerically, we
give predictions for $\tilde{e}_R \tilde{e}_R^*$ production and
compare the resummed cross section with the perturbative result.
The dependence on unphysical scales is found to be reduced, and
non-perturbative contributions remain small.
\end{abstract}
 \maketitle


\section{Introduction}\label{sec:intro}

\vspace*{-12.0cm}
\noindent{KA-TP-15-2007}\\
\noindent{LPSC 07-72} \\
\noindent{SFB-CPP-07-26}\vspace*{11.0cm}

One of the main tasks in the experimental programme of the CERN
Large Hadron Collider (LHC) is to perform an extensive and
conclusive search of the supersymmetric (SUSY) partners of the
Standard Model (SM) particles predicted by the Minimal
Supersymmetric Standard Model \cite{Nilles:1983ge, Haber:1984rc}.
Scalar leptons are among the lightest supersymmetric particles in
many SUSY-breaking scenarios \cite{Allanach:2002nj,
Aguilar-Saavedra:2005pw}. Presently, the experimental (lower)
limits on electron, muon, and tau slepton masses are 73 GeV, 94
GeV, and 81.9 GeV, respectively \cite{Yao:2006px}. Since sleptons
often decay into the corresponding SM partner and the lightest
stable SUSY particle, the distinctive signature at hadron
colliders will consist in a highly energetic lepton pair and
associated missing energy.

The leading order (LO) cross section for the production of
non-mixing slepton pairs has been calculated in
\cite{Dawson:1983fw, Chiappetta:1985ku, delAguila:1990yw,
Baer:1993ew}, while the mixing between the interaction eigenstates
was included in \cite{Bozzi:2004qq}. The next-to-leading order
(NLO) QCD corrections have been calculated in \cite{Baer:1997nh},
and the full SUSY-QCD corrections with non-mixing squarks in the
loops have been added in \cite{Beenakker:1999xh}. Recently, an
accurate calculation of the transverse-momentum ($q_T$) spectrum
including soft-gluon resummation at the next-to-leading
logarithmic (NLL) accuracy has been performed \cite{Bozzi:2006fw},
allowing for the reconstruction of the mass and the determination
of the spin of the produced particles by means of the Cambridge
(s)transverse mass variable \cite{Lester:1999tx, Barr:2005dz} and
for distinguishing thus the SUSY signal from the SM background,
mainly due to $WW$ and $t \bar t$ production \cite{Lytken:22,
Andreev:2004qq}. Very recently, the mixing effects relevant for
the squarks appearing in the loops have been investigated at NLO,
and the threshold-enhanced contributions have been computed at NLL
\cite{Bozzi:2007qr}. The numerical results show a stabilization of
the perturbative results through a considerable reduction of the
scale dependence and a modest increase with respect to the NLO
cross section.

Since the dynamical origin of the enhanced contributions is the
same both in transverse-momentum and threshold resummations, i.e.\
the soft-gluon emission by the initial state, it would be
desirable to have a formalism capable to handle at the same time
the soft-gluon contributions in both the delicate kinematical
regions, $q_{T} \ll M$ and $M^2\sim s$, $M$ being the slepton
pair invariant-mass and $s$ the partonic centre-of-mass
energy. This {\it joint} resummation formalism has been developed
in the last eight years \cite{Li:1998is, Laenen:2000ij}. The
exponentiation of the singular terms in the Mellin ($N$) and
impact-parameter ($b$) space has been proven, and a consistent
method to perform the inverse transforms, avoiding the Landau pole
and the singularities of the parton distribution functions, has
been introduced. Applications to prompt-photon
\cite{Laenen:2000de}, electroweak boson \cite{Kulesza:2002rh},
Higgs boson \cite{Kulesza:2003wn}, and heavy-quark pair
\cite{Banfi:2004xa} production at hadron colliders have exhibited
substantial effects of joint resummation on the differential cross
sections.

In this paper we apply the joint resummation formalism at the NLL
level to the hadroproduction of slepton pairs at the LHC, thus
completing our programme (started in Ref.\ \cite{Bozzi:2006fw} and
continued in Ref.\ \cite{Bozzi:2007qr}) of providing the first
precision calculations including soft-gluon resummation for
slepton pair production at hadron colliders. In Sec.\
\ref{sec:joint}, we briefly review the theoretical formalism of
joint resummation following Refs.\ \cite{Laenen:2000ij,
Kulesza:2002rh}. We reorganize the terms of the resummed formula
in a similar way as it was done for transverse-momentum
resummation in \cite{Bozzi:2005wk}. The inverse transforms from
the Mellin and impact-parameter spaces and the matching of the
resummed result with the fixed-order perturbative results are
discussed in Sec.\ \ref{sec:match}. Sec.\ \ref{sec:results} is
devoted to phenomenological predictions for the LHC, together with
a comparison of the three types of resummation
(transverse-momentum, threshold, and joint), showing their impact
on the $q_{T}$-spectrum and on the invariant-mass distribution.
Our results are summarized in Sec.\ \ref{sec:conclusions}.

\section{Joint resummation at the next-to-leading logarithmic
 order}\label{sec:joint}

We consider the hard scattering process \bea h_a(p_a) \, h_b(p_b)
\to F(M, q_T) + X,~\eea where $F$ is a generic system of colourless
particles, such as a Higgs boson or a Drell-Yan (s)lepton pair,
$M$ is the invariant mass of the final state $F$, and $q_T$ is its
transverse momentum. Thanks to the QCD factorization theorem, the
unpolarized hadronic cross section \bea\label{eq:had}
\frac{\d^2\sigma}{\d M^2\, \d q^2_T} = \sum_{a,b}\int_{\tau}^1
{\rm d}x_a \int_{\tau/x_a}^1 {\rm d}x_b\, f_{a/h_a}(x_a;\mu_F)\,
f_{b/h_b}(x_b;\mu_F) \frac{\d^2\hat{\sigma}_{ab}}{\d M^2\, \d
q^2_T} \lr z;\alpha_s, \mu_R, \mu_F \rr \eea can be written as the
convolution of the relevant partonic cross section
$\hat{\sigma}_{ab}$ with the universal distribution functions
$f_{a,b/ h_{a,b}}$ of partons $a,b$ inside the hadrons $h_{a,b}$,
which depend on the longitudinal momentum fractions of the two
partons $x_{a,b}$ and on the unphysical factorization scale
$\mu_F$. The partonic scattering cross section depends on the
strong coupling constant $\alpha_s$, the
unphysical renormalization and factorization scales $\mu_R$ and
$\mu_F$, and on the scaling variable $z=M^2/s$, where $s=x_ax_bS$
and $S=(p_a+p_b)^2$ are the partonic and hadronic centre-of-mass
energies, respectively. The lower limits for the integration over
the longitudinal momentum fractions contain the quantity
$\tau=M^2/S$, which approaches the value $\tau=1$ when the process
is close to the hadronic threshold $M^{2}\sim S$. In Mellin
$N$-space, the hadronic cross section naturally factorizes \bea
\frac{\d^2\sigma}{\d M^2\, \d q^2_T} &=& \sum_{a,b} \oint_{\cal
C}\, \frac{\d N}{2 \pi i}\, \tau^{-N}\, f_{a/h_a}(N+1; \mu_F)\,
f_{b/h_b}(N+1; \mu_F)\, \frac{\d^2\hat\sigma_{ab}}{\d M^2 \d
q_T^2}(N; \alpha_s, \mu_R, \mu_F),~\eea where the contour ${\cal
C}$ in the complex $N$-space will be specified in Sec.\
\ref{sec:match} and the $N$-moments of the various quantities are
defined according to the Mellin transform \bea F(N) = \int_0^1 \d
x\, x^{N-1}\, F(x)~\eea for $x=x_{a,b},z,\tau$ and
$F=f_{a/h_a,b/h_b}, \hat{\sigma}, \sigma$, respectively.
The jointly resummed hadronic cross section in $N$-space can be
written at NLL accuracy as \cite{Laenen:2000ij,
Kulesza:2002rh, Kulesza:2003wn} \bea \frac{\d^2\sigma^{{\rm
(res)}}}{\d M^2 \d q_T^2}(N; \alpha_s, \mu_R, \mu_F) &=& \sum_c
\hat{\sigma}_{c \bar{c}}^{(0)}\, H_{c\bar{c}}(\as, \mu_R) \int \frac{\d^2
{\bf b}}{4\, \pi} e^{i {\bf b} \cdot {\bf q}_T}\,
{\mathcal C}_{c/h_a}(N, b; \alpha_s, \mu_R, \mu_F) \nn\\
&\times& \exp\left[E_c^{\rm (PT)}(N,b;\alpha_s, \mu_R)\right]\,
{\mathcal C}_{\bar{c}/h_b}(N, b; \alpha_s, \mu_R, \mu_F).~
\label{eq:ksv}\eea The indices $c$ and $\bar c$ refer to the
initial state of the lowest-order cross section
$\hat{\sigma}^{(0)}_{c \bar{c}}$ and can then only be $q\bar{q}$
or $gg$, since the final state $F$ is assumed to be colourless.

For slepton pair and slepton-sneutrino associated production at
hadron colliders,
\bea
 h_a(p_a) \, h_b(p_b) \to  \tilde{l}_i(p_1)
 \tilde{l}_j^{(\prime)\ast}(p_2) + X,~\eea
we have $M^2=(p_1+p_2)^2$, $q_T^2=(p_{1T}-p_{2T})^2$, and
\bea
 \hat\sigma^{(0)}_{q\bar{q}} &=& \frac{\alpha^2\, \pi\,
 \beta^3}{9\,M^2} \Bigg[e_q^2\, e_l^2\, \delta_{ij} +
 \frac{e_q\,e_l\,\delta_{ij} (L_{q q Z} + R_{q q Z})\, {\rm
 Re}(L_{\tilde{l}_i \tilde{l}_j Z} + R_{\tilde{l}_i
 \tilde{l}_j Z})}{4\, x_W\,(1-x_W)\, (1-m_Z^2/M^2)}
 \nonumber\\ &+& \,\frac{(L_{q q Z}^2 + R_{q q Z}^2)
\left|L_{\tilde{l}_i \tilde{l}_j Z} + R_{\tilde{l}_i \tilde{l}_j
Z}\right|^2}{32\,x_W^2\, (1-x_W)^2
(1-m_Z^2/M^2)^2}\Bigg],~\label{eq:sig0Z}\\
\hat\sigma^{(0)}_{q\bar{q}^\prime} &=& \frac{\alpha^2\, \pi\,
\beta^3}{9\, M^2} \Bigg[\frac{\left| L_{q q^\prime W}
L_{\tilde{l}_i \tilde{\nu}_l W} \right|^2}{32\, x_W^2\, (1-x_W)^2
(1-m_W^2/M^2)^2}\Bigg]\label{eq:sig0W},~
\eea
where $i,j$ denote
slepton/sneutrino mass eigenstates with masses $m_{i,j}$, $m_Z$
and $m_W$ are the masses of the electroweak gauge bosons, $\alpha$
is the electromagnetic fine structure constant, $x_W =
\sin^2\theta_W$ is the squared sine of the electroweak mixing
angle, and the velocity $\beta$ is defined as \bea \beta &=&
\sqrt{1 + m_i^4/M^4 + m_j^4/M^4 - 2(m_i^2/M^2 + m_j^2/M^2 +
m_i^2\,m_j^2/M^4)}.~\eea The coupling strengths of the left- and
right-handed (s)fermions to the electroweak vector bosons are
given by
\bea \{ L_{f f^\prime Z}, R_{f f^\prime Z} \} &=&
(2\,T^{3}_f - 2\,e_f\,x_W) \times \delta_{f f^\prime},~\nonumber\\ \{
L_{\tilde{f}_i \tilde{f}_j^\prime Z}, R_{\tilde{f}_i
\tilde{f}_j^\prime Z} \} &=& \{ L_{f f^\prime Z}\,
S^{\tilde{f}}_{j1}\, S^{\tilde{f}^\prime\ast}_{i1}, R_{f f^\prime
Z}\, S^{\tilde{f}}_{j2}\, S_{i2}^{\tilde{f}^\prime\ast} \},~\nonumber\\
\{L_{q q^\prime W}, R_{q q^\prime W}\} &=& \{\sqrt{2}\,c_W\, V_{q
q^\prime}, 0\},~\nonumber\\ \{L_{\tilde{l}_i \tilde{\nu}_l W},
R_{\tilde{l}_i \tilde{\nu}_l W}\}
&=&\{\sqrt{2}\,c_W\,S^{\tilde{l}\ast}_{i1} ,\, 0\},~\nonumber\\
\{L_{\tilde{q}_i \tilde{q}_j^\prime W}, R_{\tilde{q}_i
\tilde{q}^\prime_l W}\} &=& \{L_{q q^\prime W}
S^{\tilde{q}\ast}_{i1}\, S^{\tilde{q}^\prime}_{j1} ,\, 0\},~
\eea
where the weak isospin quantum numbers are $T^{3}_{f}=\pm 1/2$ for
left-handed and $T^{3}_{f}=0$ for right-handed (s)fermions,
$c_{W}$ is the cosine of the electroweak mixing angle, and $V_{ff'}$ are
the CKM-matrix elements. The unitary matrices $S^{\tilde{f}}$
diagonalize the sfermion mass matrices, since in general the
sfermion interaction eigenstates are not identical to the sfermion
mass eigenstates (see App.\ \ref{sec:appA}).

The function $H_{c\bar{c}}$ in Eq.\ (\ref{eq:ksv}) contains the hard virtual
contributions and can be expanded perturbatively in powers of $\alpha_{s}$,
\bea
 H_{c\bar{c}}(\alpha_s, \mu_R) = 1 + \sum_{n=1}^{\infty}\lr
 \frac{\alpha_{s}(\mu_R)}{\pi}\rr^{n}\, H_{c\bar{c}}^{(n)}(\mu_R).~
\eea

The coefficients
\bea
 {\mathcal C}_{c/h_a}(N, b; \alpha_s,
 \mu_R, \mu_F) &=& \sum_{a,b} C_{c/b}(N; \alpha_s(M/\chi))\,
 U_{b/a}(N; M/\chi, \mu_F)\, f_{a/h_a}(N+1; \mu_F)~\label{eq:ccal}
\eea
and ${\mathcal C}_{\bar{c}/h_b}$, defined analogously, allow to evolve the
parton distribution functions $f_{a,b/h_{a,b}}$ from the unphysical
factorization scale $\mu_F$ to the physical scale $M/\chi$ with the help of
the QCD evolution operator
\bea
 U_{b/a}(N;\mu,\mu_0) = \exp \left[ \int_{\mu_0^2}^{\mu^2}
 \frac{\d q^2}{q^2} \gamma_{b/a}(N;\as(q)) \right]~
\eea
and to include, at this scale, the fixed-order contributions
\bea
 C_{c/b}(N; \alpha_s) = \delta_{cb} + \sum_{n=1}^{\infty}\lr
 \frac{\alpha_{s}}{\pi}\rr^{n}\, C_{c/b}^{(n)}(N),~
\eea
that become singular when $q_T\to0$ (but not when $z\to1$).
The QCD evolution operator fulfils the differential equation
\bea
 \frac{\d U_{b/a}(N; \mu, \mu_0)}{\d\ln\mu^2} = \sum_c
 U_{b/c}(N; \mu, \mu_0) \, \gamma_{c/a}(N;\as(\mu)),~
\eea
where the anomalous dimensions $\gamma_{c/a}(N;\as)$ are the $N$-moments of
the Altarelli-Parisi splitting functions.
The function
\bea
 \chi(\bbar, \nbar)&=&\bbar + \frac{\nbar}{1+\eta\,\bbar/\nbar}
 ~~~{\rm with}~~~ \bbar \equiv b\, M\,e^{\gamma_E}/2
 ~~~{\rm and}~~~ \nbar \equiv N e^{\gamma_E}
 \label{eq:chi}
\eea
organizes the logarithms of $b$ and $N$ in joint resummation. Its exact form
is constrained by the requirement that the leading and next-to-leading
logarithms in $\bbar$ and $\nbar$ are correctly reproduced in the limits
$\bar{b}\to \infty$ and $\bar{N}\to\infty$, respectively. The choice of
Eq.\ (\ref{eq:chi}) with $\eta = 1/4$ avoids the introduction of sizeable
subleading terms into perturbative expansions of the resummed cross section
at a given order in $\as$, which are not present in fixed-order calculations
\cite{Kulesza:2002rh}.

The perturbative (PT) eikonal exponent
\bea
 E_c^{\rm (PT)}(N,b; \alpha_s, \mu_R)
 &=& - \int_{M^2/\chi^2}^{M^2} {\d \mu^2 \over \mu^2} \, \le
 A_c(\as(\mu)) \ln \frac{M^2}{\mu^2} + B_c(\as(\mu))
 \re~\label{eq:expo}
\eea
allows to resum soft radiation in the $A$-term, while the $B$-term accounts
for the difference between the eikonal approximation and the full partonic
cross section in the threshold region, i.e.\ the flavour-conserving collinear
contributions. In the large-$N$ limit, these coefficients are directly
connected to the leading terms in the one-loop diagonal anomalous dimension
calculated in the $\overline{{\rm MS}}$ factorization scheme
\cite{Korchemsky:1988si}
\bea
 \gamma_{c/c}(N;\as) = - A_c(\as) \ln
 \nbar - \frac{B_c(\as)}{2} + \mathcal{O}(1/N).~
 \label{eq:expgamma}
\eea
They can thus also be expressed as perturbative series in $\as$,
\bea
 A_c(\alpha_s) = \sum_{n=1}^{\infty}\lr
 \frac{\alpha_{s}}{\pi}\rr^{n}\, A^{(n)}_c ~&~ {\rm and} ~&~
 B_c(\alpha_s) = \sum_{n=1}^{\infty}\lr
 \frac{\alpha_{s}}{\pi}\rr^{n}\, B_c^{(n)}.~
\eea
Performing the integration in Eq.~(\ref{eq:expo}), we obtain the form factor
up to NLL,
\bea
 E_c^{\rm (PT)}(N,b; \alpha_s, \mu_R) &=&
 g_c^{(1)}(\lambda)\,\ln\chi + g_c^{(2)}(\lambda; \mu_R)
 \label{eq:ept}
\eea
with
\bea
g_c^{(1)}(\lambda)&=& \frac{A_c^{(1)}}{\beta_0} \frac{2\, \lambda
+ \ln\big(1 - 2\, \lambda\big)}{\lambda},~\nn\\
g_c^{(2)}(\lambda; \mu_R) &=& \frac{A_c^{(1)}\,
\beta_1}{\beta_0^3} \left[ \frac{1}{2} \ln^2 \big(1 - 2\, \lambda
\big) + \frac{2\,
\lambda + \ln\big(1 - 2\, \lambda \big)}{1 - 2\, \lambda} \right]\nn\\
&+& \left[ \frac{A_c^{(1)}}{\beta_0} \ln\frac{M^2}{\mu_R^2} -
\frac{A_c^{(2)}}{\beta_0^2} \right] \left[ \frac{2\, \lambda}{1 -
2\, \lambda}+ \ln\big(1 - 2\, \lambda \big) \right] +
\frac{B_c^{(1)}}{\beta_0}\, \ln \big(1 - 2\, \lambda
\big)\label{eq:h1} ~\eea
and $\lambda = \beta_0/\pi\, \as(\mu_R)\ln\chi$.
The first two coefficients of the QCD
$\beta$-function are \bea \beta_0 = \frac{1}{12}(11\, C_A - 4\,
T_R\, N_f) ~~{\rm and}~~ \beta_1 = \frac{1}{24}(17\, C^2_A - 10\,
T_R\, C_A\, N_f - 6\, C_F\, T_R\, N_f), \eea $N_f$ being the
number of effectively massless quark flavours and $C_F = 4/3$,
$C_A = 3$, and $T_R=1/2$ the usual QCD colour factors.

In order to explicitly factorize the dependence on the parameter
$\chi$, it is possible to reorganize the resummation of the
logarithms in analogy to the case of transverse-momentum
resummation \cite{Catani:2000vq, Bozzi:2005wk}. The hadronic
resummed cross section can then be written as \bea
\frac{\d^2\sigma^{({\rm res})}}{\d M^2\, \d q^2_T} &=& \sum_{a,b}
\oint_{\cal C}\, \frac{\d N}{2 \pi i}\, \tau^{-N}\,f_{a/h_a}(N+1;
\mu_F)\, f_{b/h_b}(N+1; \mu_F) \int_{0}^{\infty} \frac{b\, \d
b}{2} \, J_0(b\, q_T)\nn\\ &\times& \sum_c\, \mathcal{H}_{ab\to
c\bar{c}}\Big(N; \alpha_s, \mu_R, \mu_F\Big) \exp\le
\mathcal{G}_c(\ln\chi; \alpha_s, \mu_R)\re.~ \label{eq:joint}
\eea The function $\mathcal{H}_{ab\to c\bar{c}}$ does not depend
on the parameter $\chi$ and contains all the terms that are
constant in the limits $b\to\infty$ or $N\to\infty$, \bea
\mathcal{H}_{ab\to c\bar{c}}\Big(N; \alpha_s, \mu_R,
\mu_F\Big) = \hat{\sigma}_{c \bar{c}}^{(0)} \left[\delta_{ca}
\delta_{{\bar c}b} + \sum_{n=1}^{\infty}\lr
\frac{\alpha_{s}(\mu_R)}{\pi}\rr^{n}\, \mathcal{H}_{ab\to
c\bar{c}}^{(n)}\Big(N; \mu_R, \mu_F\Big)\right].~
\eea
At ${\cal O}(\alpha_s)$, the coefficient $\mathcal{H}_{ab\to
c\bar{c}}^{(1)}$ is given by
\bea
 \mathcal{H}_{ab\to c\bar{c}}^{(1)}\Big(N;\mu_R,\mu_F\Big) =
 \delta_{ca} \delta_{{\bar c}b}\, H_{c\bar{c}}^{(1)}(\mu_R) +
\delta_{ca}\, C_{{\bar c}/b}^{(1)}(N) + \delta_{{\bar c}b}\,
C_{c/a}^{(1)}(N) + \left( \delta_{ca} \gamma_{{\bar c}/b}^{(1)}(N) +
\delta_{{\bar c}b} \gamma_{c/a}^{(1)}(N) \right)
\ln\frac{M^2}{\mu_F^2}.~\eea
The $\chi$-dependence appearing in the $C$-coefficient and in the evolution
operator $U$ of Eq.\ (\ref{eq:ccal}) is factorized into the exponent
$\mathcal{G}_c$, which has the same form as $E_c^{\rm (PT)}$ defined in Eq.\
(\ref{eq:expo}) except for the substitution
\bea B_c(\as) \to {\tilde B}_c(N;\as) = B_c(\as) + 2 \beta(\as)\,
\frac{\d \ln C_{c/c}(N; \as)}{\d \ln\as} + 2\, \gamma_{c/c}(N;\as).~\eea
At NLL accuracy, Eq.\ (\ref{eq:ept}) remains almost unchanged, since only
the coefficient $g_c^{(2)}$ of Eq.\ (\ref{eq:h1}) has to be
slightly modified by
\bea
 B_c^{(1)} \to {\tilde B}_c^{(1)}(N) = B_c^{(1)} + 2 \gamma_{c/c}^{(1)}(N).~
\eea
Although the first-order coefficients $C_{a/b}^{(1)}(N)$ and
$H_{c\bar{c}}^{(1)}(\mu_R)$ are in principle resummation-scheme dependent
\cite{Catani:2000vq}, this dependence cancels in the perturbative
expression of $\mathcal{H}_{ab\to c\bar{c}}$ \cite{Bozzi:2005wk}.
In the numerical code we developed for slepton pair production, we
implement the Drell-Yan resummation scheme and take $H_{q\bar{q}}(\as,\mu_R)
\equiv 1$. The $C$-coefficients are then given by \bea
C_{q/q}^{(1)}(N) = \frac{2}{3\, N\, (N+1)} +
\frac{\pi^2-8}{3}~~{\rm and}~~ C_{q/g}^{(1)}(N) = \frac{1}{2\,
(N+1)\, (N+2)}.~\eea

\section{Inverse transform and matching with the perturbative
result}\label{sec:match}

Once resummation has been achieved in $N$- and $b$-space, inverse
transforms have to be performed in order to get back to the
physical spaces. Special attention has to be paid to the
singularities in the resummed exponent, related to the divergent
behaviour near $\chi = \exp[\pi/(2 \beta_0\as)]$, i.e. the Landau
pole of the running strong coupling, and near $\bbar = -2\nbar$
and $\bbar = -4\nbar$, where $\chi=0$ and infinity, respectively.
The integration contours of the inverse transforms in the Mellin
and impact parameter spaces must therefore avoid hitting any of
these poles.

The $b-$integration is performed by deforming the integration
contour with a diversion into the complex $b$-space
\cite{Laenen:2000de}, defining two integration branches \bea b =
(\cos\varphi \pm i \sin \varphi) t ~~~~ {\rm with} ~~ 0 \leq t
\leq \infty,~ \label{eq:branch} \eea valid under the condition that the
integrand decreases sufficiently rapidly for large values of $|b|$.
The Bessel function $J_0$ is replaced by two auxiliary functions
$h_{1,2}(z,v)$ related to the Hankel functions \bea h_1(z,v)
&\equiv& - {1\over\pi}\ \int_{-iv\pi}^{-\pi+iv\pi}\, \d\theta\,
{\rm e}^{-iz\, \sin\theta},~ \nonumber\\ h_2(z,v) &\equiv& -
{1\over\pi}\ \int^{-iv\pi}_{\pi+iv\pi}\, \d\theta\, {\rm e}^{-iz\,
\sin\theta}.~\eea Their sum is always $h_1(z,v)+h_2(z,v)=2\,
J_0(z)$, but they distinguish positive and negative phases of the
$b$-contour, being then associated with only one of the two
branches defined in Eq.\ (\ref{eq:branch}).

The inverse Mellin transform is performed following a contour
inspired by the Minimal Prescription \cite{Catani:1996yz} and the
Principal Value Resummation \cite{Contopanagos:1993yq}, where one
again defines two branches \bea N = C + z\,e^{\pm i\phi}~~ {\rm
with} ~~ 0 \leq z \leq \infty,~~ \pi > \phi > \frac{\pi}{2}.~\eea
The parameter $C$ is chosen in such a way that all the
singularities related to the $N$-moments of the parton densities
are to the left of the integration contour. It has to lie within
the range $0 < C < \exp[\pi/(2 \beta_0\as)-\gamma_E]$ in order to
obtain convergent inverse transform integrals for any choice of
$\phi$ and $\varphi$.

A matching procedure of the NLL resummed cross section to the NLO
result has to be performed in order to keep the full information
contained in the fixed-order calculation and to avoid possible
double-counting of the logarithmically enhanced contributions. A
correct matching is achieved through the formula \bea
\frac{\d^2\sigma}{\d M^2\,\d q_T^2} = \frac{\d^2\sigma^{({\rm
F.O.})}}{\d M^2\,\d q_T^2}(\as) + \oint_{C_N} \frac{\d N}{2\pi
i}\, \tau^{-N} \int \frac{b\, \d b}{2} J_0(b\, q_T)
\left[\frac{\d^2\sigma^{{\rm (res)}}}{\d M^2\,\d q_T^2}(N, b; \as)
- \frac{{\rm d}^2\sigma^{{\rm (exp)}}}{\d M^2\,\d q_T^2}(N, b;
\as) \right],~ \eea where $\d^2\sigma^{({\rm F.O.})}$ is the
fixed-order perturbative result, $\d^2\sigma^ {({\rm res})}$ is
the resummed cross section discussed above, and $\d^2\sigma^{({\rm exp})}$
is the truncation of the resummed cross section to the same perturbative
order as $\d^2\sigma^{({\rm F.O.})}$. Here, we have removed the
scale dependences for brevity.

At NLO, the double-differential partonic cross section
\bea \label{eq:cross} \frac{\d\hat{\sigma}_{ab}^{({\rm F.O.})}}{\d
M^2\, \d q^2_T} (z;\alpha_s, \mu_R) = \delta(q^{2}_{T})\,
\delta(1-z)\, \hat\sigma^{(0)}_{ab} +
\frac{\alpha_s(\mu_R)}{\pi}\, \hat\sigma^{(1)}_{ab}(z) + {\cal
O}(\alpha_{s}^{2})~\eea
receives contributions from the emission of an extra gluon jet and from
processes with an initial gluon splitting into a $q\bar q$ pair,
\bea
\hat\sigma^{(1)}_{qg} (z) &=& \frac{T_R}{2\, s}
A_{qg}(s,t,u)\, \sigma^{(0)}_{q\bar{q}^{(\prime)}},~\\
\hat\sigma^{(1)}_{g\bar{q}} (z) &=& \frac{T_R}{2\, s}
A_{qg}(s,u,t)\, \sigma^{(0)}_{q\bar{q}^{(\prime)}},~\\
\hat\sigma^{(1)}_{q\bar{q}^{(\prime)}} (z) &=& \frac{C_F}{2\, s}
A_{qq}(s,t,u)\, \sigma^{(0)}_{q\bar{q}^{(\prime)}}(M)~ \eea
with \cite{Gonsalves:1989ar}
\bea
A_{qg}(s,t,u) &=& -\lr
\frac{s}{t} + \frac{t}{s} + \frac{2u\,M^2}{s t} \rr,~\\
A_{qq}(s,t,u) &=& -A_{qg}(u,t,s).~\eea
The Mandelstam variables $s$, $t$, and $u$ refer to the $2\to2$ scattering
process $ab\to\gamma,~Z^0,~W^\pm+X$ and are related to the invariant mass
$M$ (or scaled squared invariant mass $z=M^2/s$), transverse momentum $q_T$,
and rapidity $y$ of the slepton pair by the well-known relations
\bea
 s &=& x_{a}x_{b}S = M^2/z,\\ t &=&
 M^2 - \sqrt{S(M^2+q^2_T)}x_be^y,\\ u &=& M^2 -
\sqrt{S(M^2+q^2_T)}x_ae^{-y}. \eea
Integration over $q_T$ requires the cancellation of soft and collinear
singularities with virtual contributions in order to arrive at the finite
single-differential partonic cross section
\bea
 \label{eq:mass} \frac{\d\hat{\sigma}_{ab}^{({\rm F.O.})}}
 {\d M^2}(z;\alpha_s,
 \mu_R, \mu_F) = \hat\sigma^{(0)}_{ab}\,\delta(1-z) +
 \frac{\alpha_s(\mu_R)}{\pi}\, \hat\sigma^{(1)}_{ab}(z;\mu_R, \mu_F) +
 {\cal O}(\alpha_{s}^{2}),~
\eea
where the first term
$\hat\sigma^{(0)}_{ab}$ is defined in Eqs.\ (\ref{eq:sig0Z}) and
(\ref{eq:sig0W}) and the second term including the full NLO SUSY-QCD
corrections can be found in Ref.\ \cite{Bozzi:2007qr}.

The expansion of the resummed result reads \bea
\frac{\d^2\sigma^{({\rm exp})}}{\d M^2\, \d q^2_T}(N,b; \as,
\mu_R, \mu_F) = \sum_{a,b} f_{a/h_a}(N+1; \mu_F)\, f_{b/h_b}(N+1;
\mu_F)\, \hat{\sigma}^{{\rm (exp)}}_{ab}(N,b; \alpha_s, \mu_R,
\mu_F),~\eea where $\hat{\sigma}_{ab}^{{\rm (exp)}}$ is obtained
by perturbatively expanding the resummed component \bea
\hat{\sigma}^{{\rm (exp)}}_{ab}(N,b; \alpha_s, \mu_R, \mu_F) &=&
\sum_c \hat{\sigma}_{c{\bar c}}^{(0)} \Bigg\{ \delta_{ca}
\delta_{{\bar c}b} +
\sum_{n=1}^{\infty}\left(\frac{\as(\mu_R)}{\pi} \right)^n
\Bigg[{\tilde \Sigma}_{ab\to c{\bar c}}^{(n)}\left(N, \ln\chi;
\mu_R, \mu_F\right)\nn\\ &+& \mathcal{H}_{ab\to
c\bar{c}}^{(n)}\Big(N; \mu_R, \mu_F\Big)\Bigg]\Bigg\}.~\eea The
perturbative coefficients ${\tilde \Sigma}^{(n)}$ are polynomials
of degree $2n$ in $\ln\chi$, and $\mathcal{H}^{(n)}$ embodies the
constant part of the resummed cross section in the limits
$b\to\infty$ and $N\to\infty$. In particular, the first-order
coefficient ${\tilde
\Sigma}^{(1)}$ is given by \bea {\tilde \Sigma}_{ab\to c{\bar
c}}^{(1)}\left(N,\ln\chi\right) = {\tilde \Sigma}_{ab\to c{\bar
c}}^{(1;2)} \ln^2\chi + {\tilde \Sigma}_{ab\to c{\bar
c}}^{(1;1)}(N) \ln\chi,~\eea with \bea {\tilde \Sigma}_{ab\to
c{\bar c}}^{(1;2)} = - 2\, A^{(1)}_c \delta_{ca} \delta_{{\bar
c}b} ~~~~{\rm and}~~~~ {\tilde \Sigma}_{ab\to c{\bar
c}}^{(1;1)}(N) = -2\, \big(B^{(1)}_c \delta_{ca} \delta_{{\bar
c}b} + \delta_{ca} \gamma_{{\bar c}/b}^{(1)}(N) + \delta_{{\bar c}b}
\gamma_{c/a}^{(1)}(N)\Big).~\eea

\section{Numerical results}\label{sec:results}

We now present numerical results for the production of a
right-handed selectron pair at the LHC for a centre-of-mass
energy of $\sqrt{S} = 14$ TeV. For the masses and widths of
the electroweak gauge bosons and the mass of the top quark, we use
the values $m_Z = 91.1876$ GeV, $m_W = 80.403$ GeV, $\Gamma_Z =
2.4952$ GeV, $\Gamma_W = 2.141$ GeV, and $m_t = 174.2$ GeV
\cite{Yao:2006px}. The electromagnetic fine structure constant
$\alpha= \sqrt{2} G_F m_W^2\sin^2\theta_W/\pi$ is calculated in
the improved Born approximation using the world average value
$G_F=1.16637\cdot 10^{-5}$ GeV$^{-2}$ for Fermi's coupling
constant, and $\sin^2\theta_W=1- m_W^2/m_Z^2$.

We choose the mSUGRA benchmark point BFHK B \cite{Bozzi:2007me},
which gives after the renormalization group evolution of the
SUSY-breaking parameters \cite{Porod:2003um} a light ${\tilde
e}_{R}$ of mass $m_{{\tilde e}_{R}}=186$ GeV and rather heavy
squarks with masses around 800-850 GeV. The top-squark mass
eigenstate $\tilde{t}_1$ is slightly lighter, but does not
contribute to the virtual squark loops due to the negligible
top-quark density in the proton. For the LO (NLO and NLL)
predictions, we use the LO 2001 \cite{Martin:2002dr} (NLO 2004
\cite{Martin:2004ir}) MRST sets of parton distribution functions.
For the NLO and NLL predictions, $\alpha_s$ is evaluated with the
corresponding value of $\Lambda_{\overline{\rm MS}}^{n_f=5}=255$
MeV at two-loop accuracy. We allow the unphysical scales $\mu_{F}$
and $\mu_{R}$ to vary between $M/2$ and $2M$ to estimate the
perturbative uncertainty.

\begin{figure}
\centering
\includegraphics[width=.7\columnwidth]{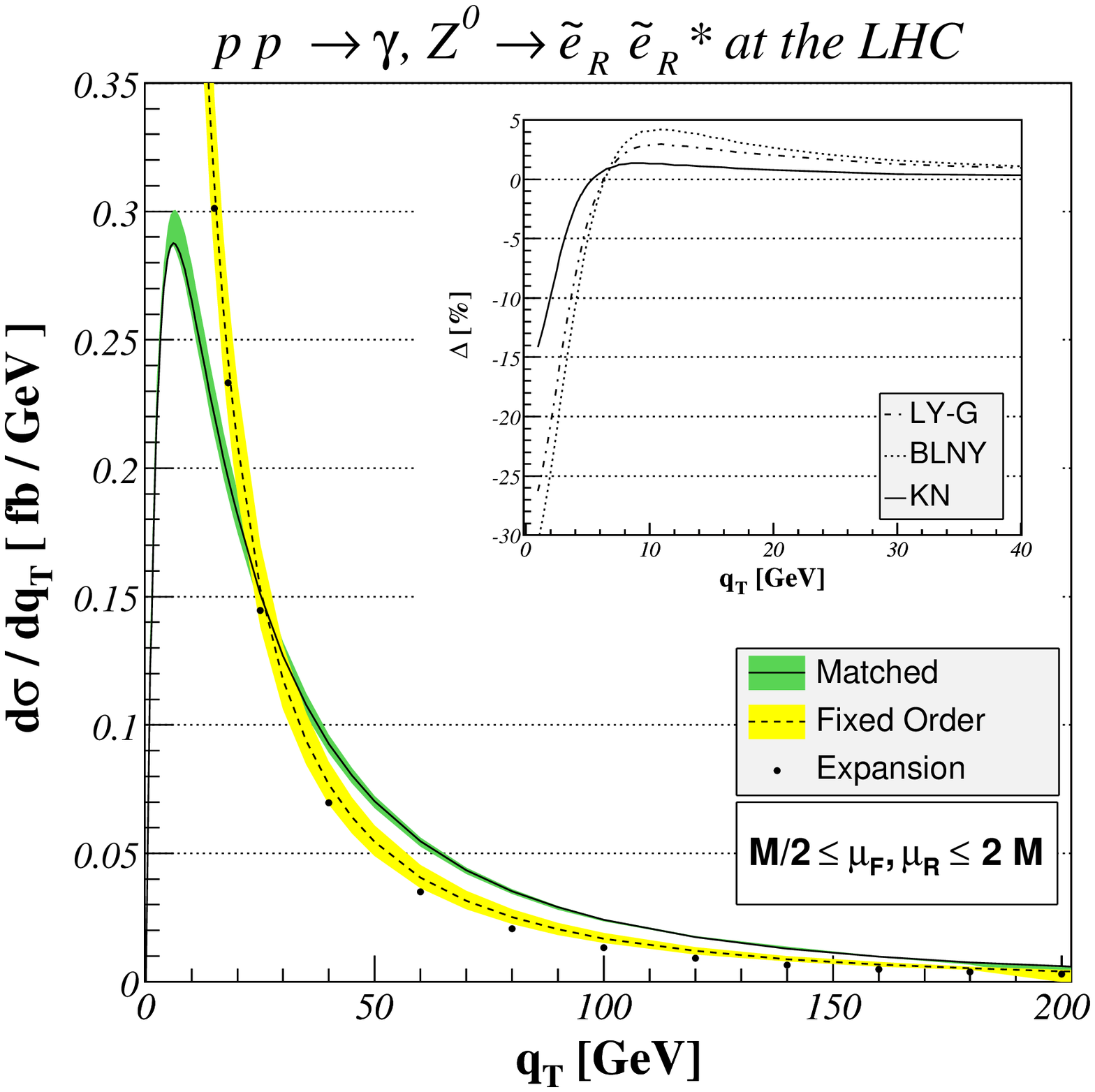}
\caption{\label{fig:1}Transverse-momentum distribution for the
process $p p \to \tilde{e}_R\, \tilde{e}_R^\ast$ at the LHC.
NLL+LO matched (full), fixed order (dotted) and asymptotically
expanded results are shown, together with three different
parameterizations of non-perturbative effects (insert).}
\includegraphics[width=.7\columnwidth]{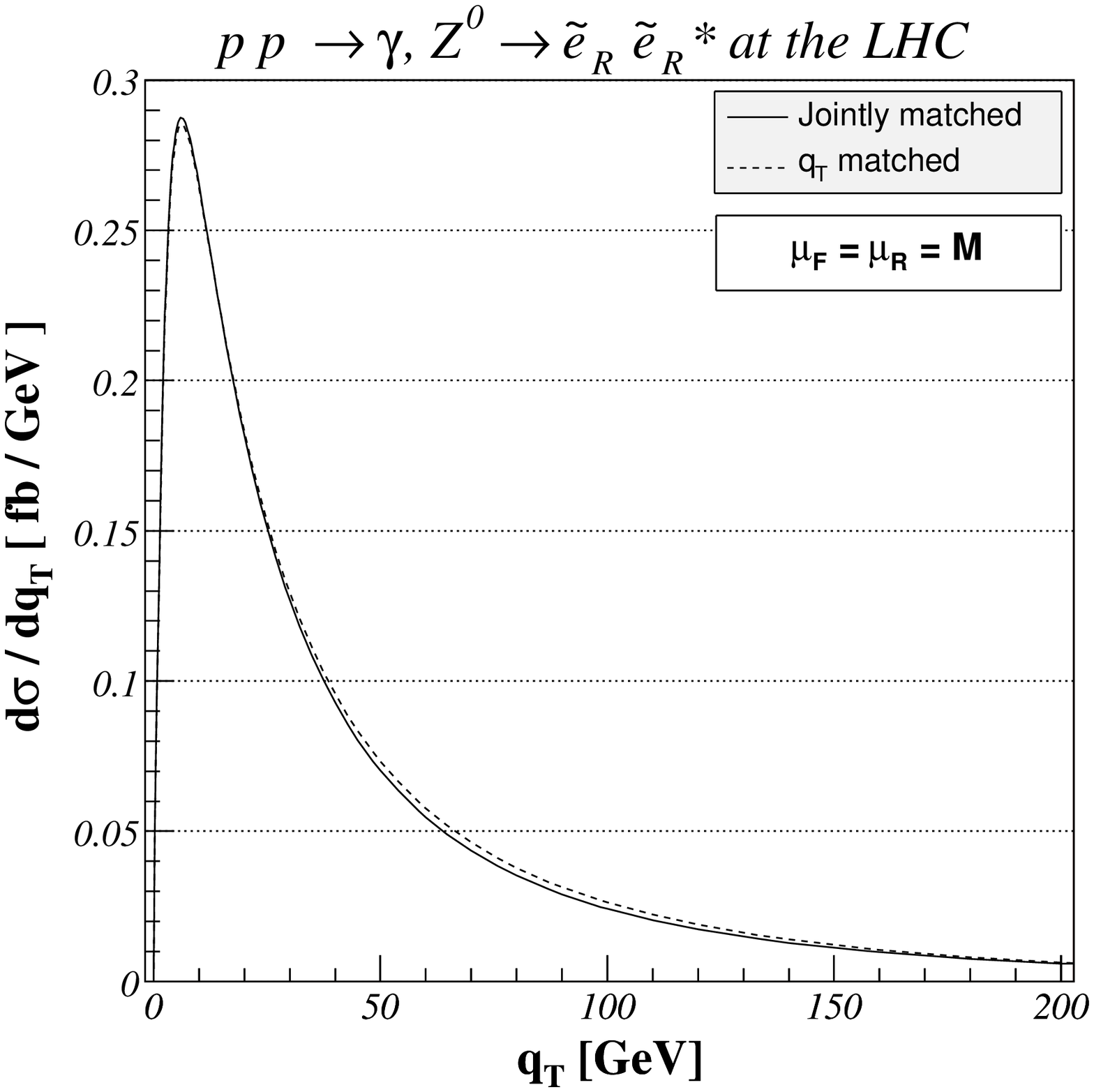}
\caption{\label{fig:2}Transverse-momentum distribution of
selectron pairs at the LHC in the framework of joint (full) and
$q_T$ (dotted) resummation.}
\end{figure}

In Fig.\ \ref{fig:1}, we present the transverse-momentum spectrum
of the selectron pair, obtained after integrating the equations of
Sec.\ \ref{sec:joint} and Sec.\ \ref{sec:match} over $M^2$, from
the $\tilde{e}_R\, \tilde{e}^\ast_R$ production threshold up to
the hadronic centre-of-mass energy. We plot the fixed order result
at order $\as$ (dashed line), the expansion of the resummed
formula at the same perturbative order (dotted line), the total
NLL+LO matched result (solid line), and the uncertainty bands from
the scale variation. The fixed order result diverges as expected
as $q_T$ tends to zero. The asymptotic expansion of the
resummation formula is in good agreement with the
$\mathcal{O}(\as)$ result in this kinematical region, since the
cross section is dominated by the large logarithms that we are
resumming. For intermediate values of $q_T$, we can see that the
agreement between the expansion and the perturbative result is
slightly worse. This effect was not present in $q_T$-resummation
for such $q_T$-values \cite{Bozzi:2006fw} and is thus related to
the threshold-enhanced contributions important in the large-$M$
region. This can also be seen in Fig.\ \ref{fig:2}, where we
directly compare the jointly- and $q_T$-matched results, the
latter having been obtained with the $q_T$-resummation formalism
of Ref.\ \cite{Bozzi:2005wk}. The two approaches lead to a similar
behaviour in the small-$q_T$ region, but the jointly-resummed
cross section is about 5\%-10\% lower than the $q_T$-resummed
cross section for transverse momenta in the range 50 GeV $< q_T <
$ 100 GeV. However, the effect of the resummation is clearly
visible in both cases, the resummation-improved result being even
40\% higher than the fixed-order result at $q_T=80$ GeV. In Fig.\
\ref{fig:1}, we also estimate the theoretical uncertainties
through an independent variation of the factorization and
renormalization scales between $M/2$ and $2M$ and show that the
use of resummation leads to a clear improvement with respect to the
fixed-order calculation. In the small and intermediate
$q_T$-regions the scale variation amounts to 10\% for the
fixed-order result, while it is always less than 5\% for the
matched result.

The $q_T$-distribution is affected by non-perturbative effects in
the small $q_T$-region coming, for instance, from partons with a
non-zero intrinsic transverse-momentum inside the hadron and from
unresolved gluons with very small transverse momentum. Global fits
of experimental Drell-Yan data allow for different
parameterizations of these effects, which can be consistently
included in the resummation formula of Eq.\ (\ref{eq:joint})
through a non-perturbative form factor $F_{ab}^{{\rm NP}}$. We
include in our analysis three different parameterizations of this
factor \cite{Ladinsky:1993zn, Landry:2002ix, Konychev:2005iy},
\bea \label{eq:NP1}F_{ab}^{{\rm NP}(LY-G)}(b,M,x_1,x_2) &=&
\exp\left[-b^2\left(\bar{g}_1 \!+\! \bar{g}_2 \ln\frac{b_{{\rm
max}}M}{2} \right) \!-\! b\, \bar{g}_1 \,\bar{g}_3\, \ln(100\,x_1
x_2) \right],~~~\\\label{eq:NP2} F_{ab}^{{\rm
NP}(BLNY)}(b,M,x_1,x_2) &=& \exp \left[-b^2\left(\tilde{g}_1 +
\tilde{g}_2 \ln\frac{b_{{\rm max}}M}{2} + \tilde{g}_1\,
\tilde{g}_3 \ln(100\,x_1 x_2) \right) \right],~~~\\
\label{eq:NP3} F_{ab}^{{\rm NP}(KN)}(b,M,x_1,x_2) &=& \exp
\left[-b^2\left(a_1 + a_2 \ln\frac{M}{3.2 {\rm GeV}} + a_3
\ln(100\, x_1\, x_2)\right)\right].~~~\eea The most recent values
for the free parameters in these functions can be found in Refs.\
\cite{Landry:2002ix,Konychev:2005iy}. We show in the upper-right
part of Fig.\
\ref{fig:1} the quantity \bea \Delta = \frac{d\sigma^{\rm
(res.+NP)}(\mu_R=\mu_F=M)-d\sigma^{(\rm
res.)}(\mu_R=\mu_F=M)}{d\sigma^{(\rm res.)} (\mu_R=\mu_F=M)},\eea
which gives thus an estimate of the contributions from the
different NP parameterizations (LY-G, BLNY and KN). They are under
good control, since they are always less than 5\% for $q_{T}>$ 5
GeV and thus considerably smaller than the resummation effects.

\begin{figure}
\centering
\includegraphics[width=.7\columnwidth]{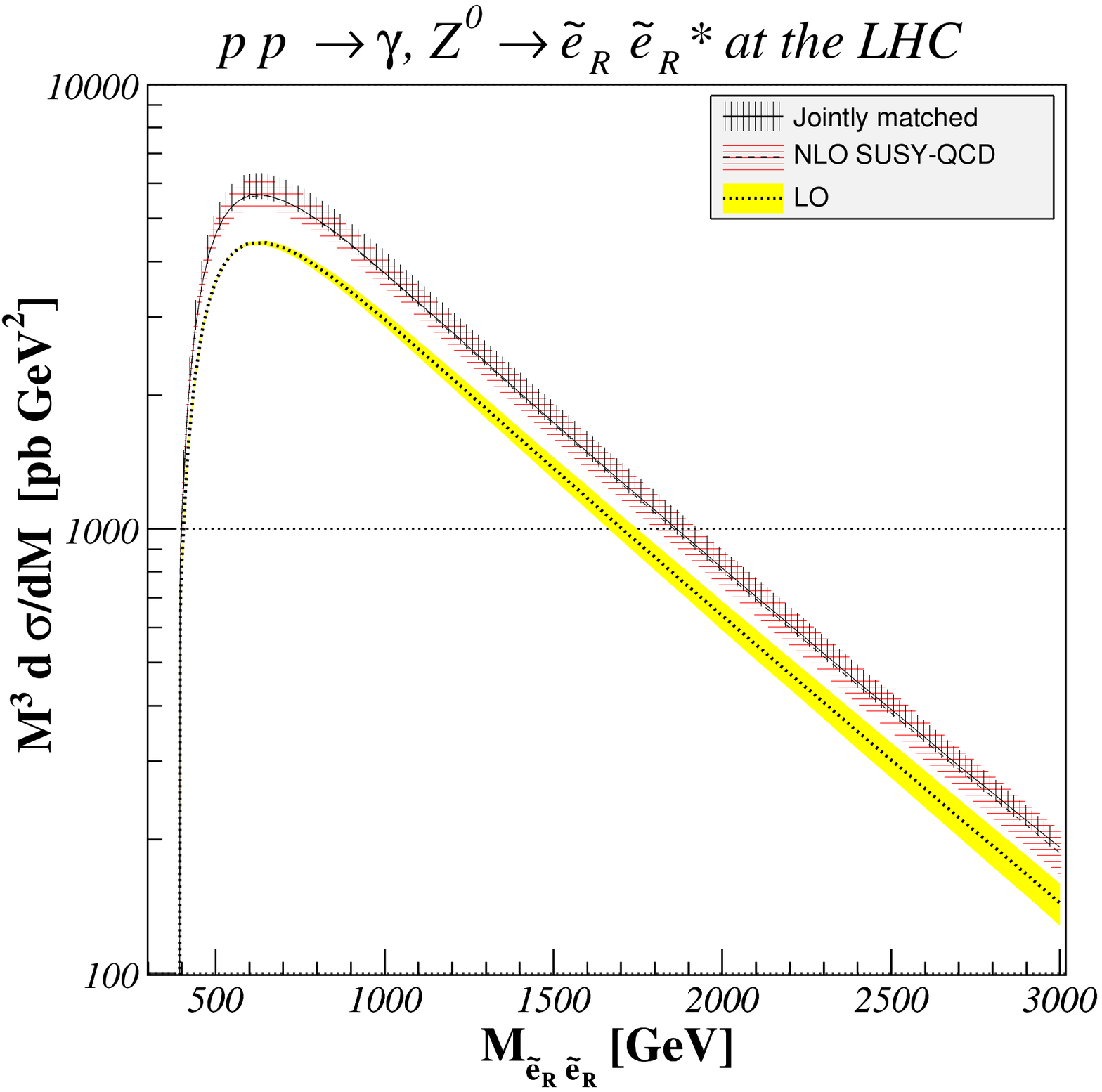}
\caption{\label{fig:3}Invariant-mass distribution
$M^3\,\d\sigma/\d M$ of $\tilde{e}_R$ pairs at the LHC. We show
the total NLL+NLO jointly matched (full), as well as the
fixed-order NLO SUSY-QCD (dashed) and LO QCD (dotted) results,
with the corresponding scale uncertainties (vertically hashed,
horizontally hashed, and shaded bands).}
\includegraphics[width=.7\columnwidth]{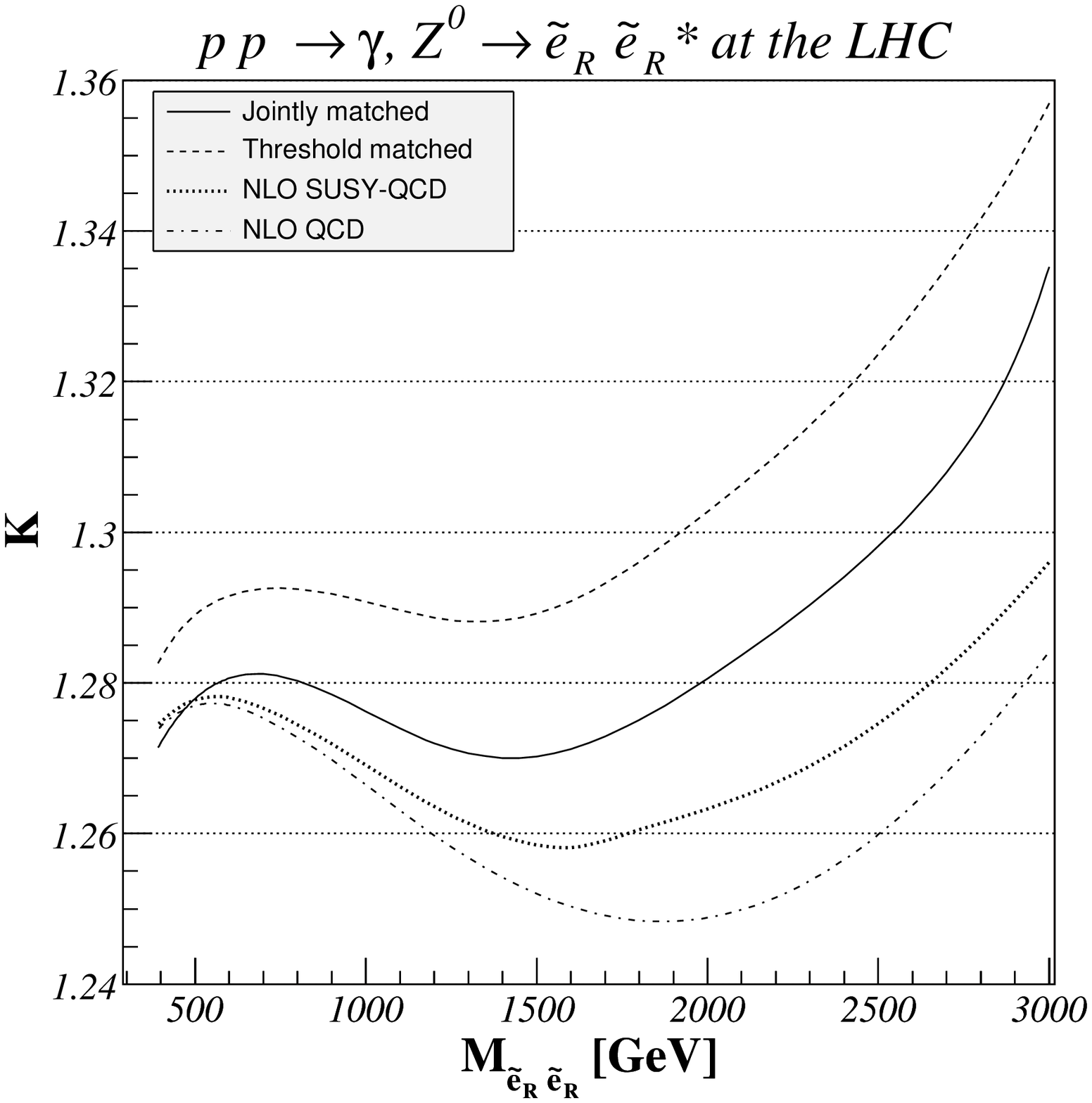}
\caption{\label{fig:4}$K$-factors as defined in Eq.\
(\ref{eq:Kbis}) for $\tilde{e}_R$ pair production at the LHC. We
show the total NLL+NLO jointly (full), and threshold (dashed)
matched results, as well as the fixed-order NLO SUSY-QCD (dotted)
and QCD (dash-dotted) results.}
\end{figure}

The invariant-mass distribution $M^3\d\sigma/\d M$ for
$\tilde{e}_R$-pair production at the LHC is obtained after integrating the
equations of Sec.\ \ref{sec:joint} and Sec.\ \ref{sec:match} over $q_T^2$
and is shown in Fig.\ \ref{fig:3}. The
differential cross section $\d\sigma/\d M$ has been multiplied by
a factor $M^3$ in order to remove the leading mass dependence of
propagator and phase space factors. We can see the $P$-wave
behaviour relative to the pair production of scalar particles, since
the invariant-mass distribution rises above the threshold at
$\sqrt{s}=2 m_{\tilde{e}_R}$ with the third power of the slepton
velocity and peaks at about 200 GeV above threshold (both for
$M^3\d\sigma/\d M$ and the not shown $\d\sigma/\d M$ differential
distribution), before falling off steeply due to the $s$-channel
propagator and the decreasing parton luminosity.
In the large-$M$ region, the resummed cross section is 30\% higher
than the leading order cross section, but this represents only a
3\% increase with respect to the NLO SUSY-QCD result. In the
small-$M$ region, much further then from the hadronic threshold,
resummation effects are rather limited, inducing a modification of
the NLO results smaller than 1\%.
The shaded, horizontally, and vertically hashed bands in Fig.\
\ref{fig:3} represent the theoretical uncertainties for the LO,
NLO SUSY-QCD, and the jointly-matched predictions. At LO the
dependence comes only from the factorization scale and increases
with the momentum-fraction $x$ of the partons in the proton (i.e.\
with $M$), being thus larger in the right part of the figure. This
dependence is largely reduced at NLO due to the factorization of
initial-state singularities in the PDFs. Including the dependence
due to the renormalization scale in the coupling
$\alpha_s(\mu_R)$, the total variation is about 7\%-11\%. After
resummation, the total scale uncertainty is finally reduced to
only 7\%-8\% for the matched result, the reduction being of course
more important in the large-$M$ region, where the resummation
effects are more important.

In Fig.\ \ref{fig:4}, we show the cross section correction factors
\bea \label{eq:Kbis} K^i = \frac{{\rm d}\sigma^i / {\rm d}M}{{\rm
d}\sigma^{\rm LO} / {\rm d}M} \eea as a function of the
invariant-mass $M$. $i$ labels the corrections induced by NLO QCD,
NLO SUSY-QCD, joint- and threshold-resummation (as obtained in
\cite{Bozzi:2007qr}), these two last calculations being matched
with the NLO SUSY-QCD result. At small invariant mass $M$, the
resummation is less important, since we are quite far from the
hadronic threshold, as shown in the left part of the plot. At
larger $M$, the logarithms become important and lead to a larger
increase of the resummed $K$-factors over the fixed-order one. We
also show the difference between threshold and joint resummations,
which is only about one or two percents. This small difference is
due to the choice of the Sudakov form factor $\mathcal G$ and of
the $\mathcal H$-function, which correctly reproduce
transverse-momentum resummation in the limit of $b\to\infty$, $N$
being fixed, but which present some differences in the pure
threshold limit $b\to 0$ and $N\to\infty$, as it was the case for
joint resummation for Higgs and electroweak boson production
\cite{Kulesza:2002rh, Kulesza:2003wn}. However, this effect is
under good control, since it is much smaller than the theoretical
scale uncertainty of about 7\%.

\section{Conclusions}\label{sec:conclusions}
With this work we complete our programme of performing precision
calculations for slepton pair production at hadron colliders.
Together with the previous papers on transverse-momentum
\cite{Bozzi:2006fw} and threshold \cite{Bozzi:2007qr} resummation,
soft-gluon resummation effects are now consistently included in
predictions for various distributions exploiting the $q_{T}$,
threshold, and joint resummation formalisms. We found that the
effects obtained from resumming the enhanced soft contributions
are important at hadron colliders, even far from the critical
kinematical regions where the resummation procedure is fully
justified. The numerical results show a considerable reduction of the
scale uncertainty with respect to fixed order results and also a
negligible dependence on non-perturbative effects, introduced
through different Gaussian-like smearings of the Sudakov form
factors. These features lead to an increased stability of the
perturbative results and thus to a possible improvement of the
slepton pair (slepton-sneutrino) search strategies at the LHC.

\acknowledgments This work was supported by a Ph.D.\ fellowship of
the French ministry for education and research.

\appendix
\section{Sfermion Mixing}
\label{sec:appA}

The soft SUSY-breaking terms $A_f$ of the trilinear
Higgs-sfermion-sfermion interaction and the off-diagonal Higgs
mass parameter $\mu$ in the MSSM Lagrangian induce mixings of the
left- and right-handed sfermion eigenstates $\tilde{f}_{L,R}$ of
the electroweak interaction into mass eigenstates
$\tilde{f}_{1,2}$. The sfermion mass matrix is given by
\cite{Haber:1984rc} \bea
 {\cal M}^2 &=& \left(\begin{array}{cc} m_{LL}^2+m_f^2 & m_f\,m_{LR}^\ast \\
 m_f\,m_{LR}& m_{RR}^2+m_f^2 \end{array}\right)
\eea with \bea
 m_{LL}^2&=&m_{\tilde{F}}^2 + (T_f^3-e_f\,\sin^2\theta_W)\,m_Z^2\,\cos2\beta,\\
 m_{RR}^2&=&m_{\tilde{F}^\prime}^2 + e_f\,\sin^2\theta_W\,m_Z^2\,\cos2\beta,\\
 m_{LR}&=&A_f-\mu^\ast\left\{
 \begin{array}{l}
 \cot\beta\hspace*{3.mm}{\rm for~up-type~sfermions.}
 \\ \tan\beta\hspace*{2.8mm}{\rm for~down-type~sfermions.}
 \end{array}\right.\hspace*{3mm}
\eea It is diagonalized by a unitary matrix $S^{\tilde{f}}$,
$S^{\tilde{f}}\, {\cal M}^2\,S^{\tilde{f}\dagger}={\rm
diag}\,(m_1^2,m_2^2)$, and has the squared mass eigenvalues \bea
 m_{1,2}^2=m_f^2+{1\over 2}\Big( m_{LL}^2 + m_{RR}^2
 \mp\sqrt{(m_{LL}^2-m_{RR}^2)^2 + 4\,m_f^2\,|m_{LR}|^2} \Big).
\eea For real values of $m_{LR}$, the sfermion mixing angle
$\theta_{\tilde{f}}$, $0 \leq \theta_{\tilde{f}} \leq \pi/2$, in
\bea
 S^{\tilde{f}} = \left( \begin{array} {cc}~~\,\cos\theta_{\tilde{f}} &
 \sin\theta_{\tilde{f}} \\
 -\sin\theta_{\tilde{f}} & \cos\theta_{\tilde{f}} \end{array} \right)
 \hspace*{1mm} {\rm with} \hspace*{1mm} \left(
 \begin{array}
 {c} \tilde{f}_1 \\
 \tilde{f}_2
 \end{array} \right) =
 S^{\tilde{f}} \left(
 \begin{array}
 {c} \tilde{f}_L \\ \tilde{f}_R
 \end{array} \right) \hspace*{5mm}&&
\eea can be obtained from
\begin{equation}
 \tan2\theta_{\tilde{f}}={2\,m_f\,m_{LR}\over m_{LL}^2-m_{RR}^2}.
\end{equation}
If $m_{LR}$ is complex, one may first choose a suitable phase
rotation $\tilde{f}_R^\prime = e^{i\phi}\tilde{f}_R$ to make the
mass matrix real and then diagonalize it for $\tilde{f}_L$ and
$\tilde{f}_R^\prime$. $\tan \beta=v_u/v_d$ is the (real) ratio of
the vacuum expectation values of the two Higgs fields, which
couple to the up-type and down-type (s)fermions. The soft
SUSY-breaking mass terms for left- and right-handed sfermions are
$m_{\tilde{F}}$ and $m_{\tilde{F}^\prime}$ respectively.


\end{document}